\documentclass[letterpaper,10pt,conference]{ieeeconf}

\IEEEoverridecommandlockouts
\usepackage{graphicx,psfrag,enumerate}
\usepackage{amssymb}
\usepackage{amsmath}
\usepackage{calrsfs}

\newtheorem{theorem}{Theorem}

\newtheorem{lemma}{Lemma}

\newtheorem{assumption}{Assumption}
\begin{document}

\title{Control over adversarial packet-dropping communication networks revisited%
\thanks{This work was supported by the Australian Research Council under
  Discovery Projects funding scheme (project DP120102152) and, in parts, by
  US Air Force Office of Scientific Research (AFOSR) under grant number
  MURI FA 9550-10-1-0573, and the US National Science Foundation under
  award \#1151076.}
\thanks{A version of this paper has been accepted for presentation at the
  2014 American Control Conference, Portland, Oregon, USA.}}

\author{V.~Ugrinovskii\thanks{School of Engineering and IT, University of NSW
  at the   Australian Defence Force Academy, Canberra, ACT, 2600, Australia,
Email: v.ugrinovskii@gmail.com} \and 
C. Langbort\thanks{Department of
Aerospace Engineering, University of Illinois at Urbana-Champaign. Email:
langbort@illinois.edu}}

\maketitle

\begin{abstract}
We revisit a one-step control problem over an adversarial packet-dropping
link. The link is modeled as a set of binary channels controlled by a
strategic jammer whose intention is to wage a
`denial of service' attack on the plant 
by choosing a most damaging channel-switching strategy.
The paper introduces a class of zero-sum games between the jammer and
controller as a scenario for such attack, and derives necessary and
sufficient conditions for these
games to have a nontrivial saddle-point equilibrium. At this equilibrium,
the jammer's optimal policy is to randomize in a region of the plant's
state space, thus requiring the controller to undertake a nontrivial
response which is different from what one would expect in a standard
stochastic control problem over a packet dropping channel.  
\end{abstract}

\section{Introduction and Motivation}\label{Intro}

The topic of control over a communication channel has been extensively
studied in the past decade, with issues such as the minimum data rate for stabilization \cite{Tak, Nair, Yuk, Min, Mart} and optimal quadratic
closed-loop performance \cite{Sche, Imer, Gar} being the main focus. Other
issues of interest concern effects of channel-induced packet
drops and/or time-varying delays on closed-loop performance.

The majority of papers concerned with control over networks regards the 
mechanism of information loss in the network as probabilistic but
not strategic. In contrast, in the problem of control over an
\emph{adversarial} channel, the communication link is controlled by a rogue
jammer whose intention is to mount a cyber attack on the system
by actively jamming the communication link. 
Its objectives are to impose on the controller a control
law which cannot be expected under regular operating conditions in a
packet-dropping network. If the controller is unaware of the jammer's
actions and continues to follow a control policy designed for a regular
network, the system performance is likely to be
inferior. It is this situation that is considered as a scenario of a
successful cyber attack by the jammer.   

A natural way to describe the problem of control over an
adversarial channel is to employ a game-theoretic
formulation. Originally proposed in~\cite{Amin}, this idea has been
followed upon in a number of recent papers including~\cite{Abhi,LaU3a}.  
A zero-sum dynamic game between a controller performing a finite horizon
linear-quadratic control task and a jammer, proposed in \cite{Abhi},
specifically accounted for the jammer's strategic intentions and
limited actuation capabilities, but was otherwise agnostic regarding the
type of channel involved. A startling conclusion of \cite{Abhi} was that
in order to maximally disrupt the control task, the jammer must act in a
markedly different way than a legitimate,  non-malicious, packet-dropping
channel. Once this deterministic behavior is observed by the controller, it
can establish with certainty that an attack has taken place.  

In~\cite{LaU3a}, we introduced a different model, which, while capturing
the same fundamental aspects of the problem as in \cite{Abhi}, modified the
jammer's action space so that each jammer decision corresponded to a choice
of  channel rather than to passing/blocking transmission. The
corresponding one-step zero-sum game was shown to have a unique saddle
point in the space of mixed jammer strategies. In turn, the controller's
best response to the jammer's 
optimal randomized strategy was to act as if it was operating over a
packet-dropping channel whose statistical characteristics were controlled
by the jammer. Since under normal circumstances the controller cannot be
aware of these characteristics, and cannot implement such a
best response strategy, we regard the zero-sum game in~\cite{LaU3a} as an
example of a successful cyber attack. 

In this paper, we show that such a situation is not specific to the
zero-sum game considered in~\cite{LaU3a}. We introduce a
class of zero-sum stochastic games that generalize the model introduced
in~\cite{LaU3a}. For these games we obtain necessary and sufficient
conditions which guarantee the existence of optimal
jammer's strategies whose nature suggests that the jammer must select
its actions randomly, in order to make a maximum impact on the control
performance. Our conditions are quite general, they apply to nonlinear
systems and draw on standard convexity/coercivity properties of payoff
functions. Furthermore, we specialize these conditions to the 
linear-quadratic control problem over a packet-dropping link considered
in~\cite{LaU3a} and show that our conditions allow for an express 
characterization of a set of plant's initial states for which optimal
randomized jammer strategies exist (this is in contrast to~\cite{LaU3a}
where a complete analysis of the state space had to be performed to
determine such regions). We also compute an optimal controller
response to those strategies, which turns out to be nonlinear. 

Our analysis is restricted to one-step zero-sum games. Although such a
formulation is admittedly 
simple, due to the general nature of the game under consideration, it can
be thought of as reflecting a more general situation where one is dealing
with a one-step Hamilton-Jacobi-Bellman-Isaacs min-max problem associated
with a multi-step optimal control problem. 
Also, even a one-step formulation provides a rich insight into
a possible scenario of cyber attacks on controller networks. We believe
that such an insight can be valuable as was the case, e.g., in early
studies of adversarial channels and multi-agent decision problems involving
incomplete information \cite{B&W}.  

We present our model in Section~\ref{sec_mod}. The problem formulation,
its assumptions and preliminary results are given in
Section~\ref{sec_form}. The main result of the paper that gives a necessary
and sufficient condition for the game under consideration to have a
nontrivial saddle point is presented in Section~\ref{main}. Next, in
Section~\ref{Examples} we demonstrate an application of this result to the
linear-quadratic static problem which is an extension of the problem
in~\cite{LaU3a}. In this problem, the 
jammer is offered an additional reward for undertaking actions concealing its
presence. Conclusions are given in Section~\ref{Conclusions}.    

\section{Model description}
\label{sec_mod}

The general model description is an extension of that in ~\cite{LaU3a}. 
We consider a situation where a strategic jammer is attacking the link in
the feedback loop between a controller and a plant. The plant is a 
general discrete-time system described by  
\begin{equation}
\label{plant}
x^+=F(x,v)
\end{equation}
with a given initial condition; $x\in \mathbb{R}^N$ is the state, $v$
is a scalar input, $F(\cdot,\cdot)$ is an $\mathbb{R}^N$-valued function
defined on 
$\mathbb{R}^N\times \mathbb{R}$.  

The plant input $v$ and the control signal $u$ are related by the equation 
\[
v=bu,
\]
which describes the transmission of information from the controller to the plant
over a packet-dropping communication link. Here, $b$ is a discrete
random variable taking value in $\{0,1\}$, that describes the transmission
state of the link. 
The value of $b$ depends on actions of the jammer and the state
of the communication link, as explained below.

\begin{figure}
\psfrag{1}{1}
\psfrag{0}{0}
\psfrag{qj}{$q_j$}
\psfrag{1-qj}{$1-q_j$}
\begin{center}
\vskip 1.5ex

\includegraphics[width=0.7\columnwidth]{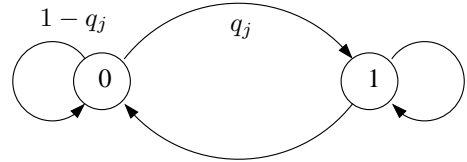}
\caption{Each channel in $\mathcal{F}$ is a binary channel. Shown here is channel $f_j$. \label{fig_2}}
\end{center}
\end{figure}

The communication link consists of a finite set $\mathcal{F}$ of channels (with
$|\mathcal{F}| =n$) out of which the jammer can draw with certain probability a
channel to replace the currently active channel so as to optimally disrupt the
control task. 
Each channel can be either in passing or blocking state, and the
transmission states of \emph{all} channels randomly change \emph{after} one
of them is selected as a replacement. Hence, each channel 
$f_j\in\mathcal{F}$ represents a binary channel with the state space $\{0,1\}$,
as pictured in Figure~\ref{fig_2}. To describe the probability model of channel
transitions, let $c^-, c \in \{0,1\}^n$ denote the vectors of
transmission states of \emph{all} channels before and after the jammer
has selected one of them to replace the current one, respectively, with the
$j$th component $c_j^-$, $c_j$ denoting the corresponding transmission
state of channel $f_j$. The probability of channel $f_j$ to become  
``passing'' after the replacement is selected, given its and all other
channels' previous transmission states, is then  
\begin{equation}
q_j = \mathrm{Pr}(c_j = 1 | c^-).
\label{q_j}
\end{equation} 

The jammer strategy is to choose a probability distribution over $\mathcal{F}$,
indicating which channel it desires to switch to. We denote this
distribution by a vector $p$ in the unit simplex $\mathcal{S}_{n-1}$ of
$\mathbb{R}^n$. That is, the jammer's strategy is to influence the
selection of a channel linking the controller to the plant. Let the index
of the selected channel be $S$, then $S$ is a discrete random variable
taking values in $\{1,...,n\}$, distributed in accordance with the
vector $p$. The latter depends on the information set available to
the jammer which includes the current state of the plant $x$,
the index of the channel occupying the link $j^-$, and the vector of
transmission states of all channels $c^-$ which jammer observes before the
link switches from channel $f_{j^-}$ to a new channel:
\begin{equation}\label{p_j}
p_j=\mathrm{Pr} (S = j| x, c^-, j^-) \mbox{ for all } j=1,...,n.
\end{equation} 
In addition, if the control input $u$ is available to the jammer, the
vector $p$ may depend on $u$ as well.

After the jammer has made its decision, the random variable $S$ is realized and 
the link switches to the channel $f_S$. After that, the transmission state
of all channels including $f_S$ changes, according to (\ref{q_j}).
Thus, the jammer cannot predict the transmission state of the channels in
$\mathcal{F}$ when selecting the probability vector $p$. 

In accordance with this channel switching mechanism, the
transmission state of the link between the controller and the plant is
determined by the binary random variable $c_S$, i.e., $b=c_S$, which takes
value 1 with probability $p'q$ and value 0 with probability $1-p'q$.  

Clearly, $b$ and $S$ are statistically dependent. All random variables
considered in this paper will be adapted to the joint 
conditional probability distribution of $S$ and $b$, given $x$, $j^-$ and
$c^-$. The expectation with respect to this conditional probability
distribution is denoted
$\mathbb{E}[\cdot]$.

\section{Problem formulation and preliminary results}
\label{sec_form}

We now introduce a general two-player stochastic one-step zero-sum game as
follows. In this game, we assume that the initial state
of the plant $x$, the initial vector of transmission states $c^-$ and the
channel that initially occupies the link $f_{j^-}$ are known to both the
jammer and controller.  Let $\sigma(y,u,f)$ be a scalar
function of $(y,u,f)\in\mathbb{R}^N\times \mathbb{R}\times
\mathcal{F}$. This function will determine the payoff 
of the game played by the controller and the jammer. 
The standing assumptions regarding this function are summarized below:
\begin{assumption}\label{sigma.A1}
For all $f_j\in\mathcal{F}$, $\sigma(\cdot,\cdot,f_j)\in
C^1(\mathbb{R}^N\times \mathbb{R})$. 
\end{assumption}

\begin{assumption}\label{sigma.A3}
For each $f_j\in\mathcal{F}$ and $x\in \mathbb{R}^n$, the functions
$\sigma(F(x,0),\cdot,f_j)$ and  $\sigma(F(x,\cdot),\cdot,f_j)$
are coercive.  
\end{assumption}

\begin{lemma}\label{U}
Under Assumptions~\ref{sigma.A1} and~\ref{sigma.A3}, for every $x\neq 0$
there exists a compact set $U(x)\subset \mathbb{R}$ with the properties:
\begin{enumerate}[(i)]
\item 
For all $f_j\in \mathcal{F}$,
\begin{eqnarray}
\lefteqn{\inf_{u\in \mathbb{R}}\max_{j}
  \mathbb{E}[\sigma(x^+,u,f_S)|S=j]} && \nonumber \\
&& = \inf_{u\in U(x)}\max_{j} \mathbb{E}[\sigma(x^+,u,f_S)|S=j)].   
\label{max.property}
\end{eqnarray} 
\item
\begin{equation}
  \label{infU.1}
  \inf_u\mathbb{E}[\sigma(x^+,u,f_S)]=\inf_{u\in
    U(x)}\mathbb{E}[\sigma(x^+,u,f_S)]. 
\end{equation}
\end{enumerate}
\end{lemma}

The proof is omitted for the sake of brevity. It proceeds by first proving that 
the coercivity of the functions involved ensures that the infima on the
left hand side of (\ref{max.property}) and (\ref{infU.1}) exist. In
particular, $\inf_u h(u)>-\infty$, $h(u)\triangleq \max_{j} h_j(u)$, $h_j(u)\triangleq \mathbb{E}[\sigma(x^+,u,f_S)|S=j]$. Next we
show that a suitably defined set $U_\alpha=\{u: h(u)\le \alpha\}$,
with a sufficiently large $\alpha> \inf_u h(u)$ can be chosen as $U(x)$.

We now define the stochastic zero-sum min-max game of
interest for the plant (\ref{plant}). 
In this game, the controller is a minimizing player who selects the
control input $u \in \mathbb{R}$ based on $x$, $j^-$ and $c^-$. Also, the
jammer is the maximizing player who chooses a probability distribution
vector $p\in\mathcal{S}_{n-1}$ for the `channel selection' random variable
$S$, as the function of $x$, $j^-$, $c^-$ and possibly $u$, as in (\ref{p_j}). 
The controller's best action is determined by computing
\begin{eqnarray}  
J_1 &=& \inf_u \max_{p \in \mathcal{S}_{n-1}}
\mathbb{E}\left[\sigma(x^+,u,f_S)\right]. \label{inf_sup} 
\end{eqnarray}
while the jammer's best action is obtained by computing  
\begin{eqnarray} 
J_2 &=& \max_{p \in \mathcal{S}_{n-1}} \inf_u
\mathbb{E}\left[\sigma(x^+,u,f_S)\right]. 
\label{sup_inf} 
\end{eqnarray}
Our goal is to show that $J_1=J_2$, i.e., that the corresponding zero-sum
game has a value.

Lemma~\ref{U} allows to reduce the minimization over $u\in\mathbb{R}$ in (\ref{inf_sup})
and (\ref{sup_inf}) to minimization over a compact set
$U(x)$. Indeed, the cost function of the inner maximization problem
(\ref{inf_sup}) is linear in $p$, therefore using claim (i) of
Lemma~\ref{U} leads to the conclusion that
\begin{eqnarray}
J_1&=&\inf_u\max_j\mathbb{E} \left[\sigma(x^+,u,f_j)\right] \nonumber \\
&=& \inf_{u\in U(x)}\max_j \mathbb{E}
\left[\sigma(x^+,u,f_j)\right] \label{inf_U_max} \\
&=& \inf_{u\in U(x)}\max_{p \in \mathcal{S}_{n-1}} \mathbb{E}
\left[\sigma(x^+,u,f_S)\right]. 
\label{inf_U_sup}
\end{eqnarray}
Also, it follows from claim (ii) of Lemma~\ref{U} that for every $p$, the inner
minimization problem in (\ref{sup_inf}) can be carried out over $U(x)$. Thus,  
\begin{eqnarray}
J_2&=& \max_{p \in \mathcal{S}_{n-1}} \inf_{u\in U(x)}
\mathbb{E}\left[\sigma(x^+,u,f_S)\right].
\label{sup_inf_U}
\end{eqnarray}

We make an additional assumption about the set $U(x)$.

\begin{assumption}\label{sigma.A4}
The set $U(x)$ is connected.
\end{assumption}

Under this assumption, the set $U(x)$ is a closed bounded interval,
hence it is a convex set.  
Of course, this can be guaranteed when
$\sigma$ is chosen so that each $h_j$ is convex.      

\begin{lemma}\label{BO.T4}
Under Assumptions~\ref{sigma.A1}, ~\ref{sigma.A3}, and \ref{sigma.A4}, 
the value of the game (\ref{inf_sup}) exists, i.e.,
$
-\infty<J_1=J_2<\infty.
$  
Furthermore, the game has
a (possibly non-unique) saddle point. 
\end{lemma}

It is not unreasonable to assume that in the game (\ref{inf_sup}) the
jammer, who observes the controller action $u$, can rank all the channels
according to the contribution they make towards the payoff and order them
accordingly. It can do so by comparing the conditional expected cost values
$h_j(u)=\mathbb{E}[\sigma(x^+,u,S)|S=j]$.  

\begin{assumption}\label{sigma.A2}
For any two channels $f_j,f_k\in\mathcal{F}$, $j,k\neq j^-$, if $j<k$ then
\begin{eqnarray}
  \label{chan.ineq}
  \mathbb{E}[\sigma(x^+,u,f_S)|S=j]\ge 
  \mathbb{E}[\sigma(x^+,u,f_SS)|S=k] \\
\nonumber \forall u\in U(x). 
\end{eqnarray}
\end{assumption}

Assumption~\ref{sigma.A2} generalizes the situation considered
in~\cite{LaU3a} where all channels were ranked according to the probability
of becoming passing, $q_1 <q_2<...<q_n$. In Section~\ref{Examples} we will
show that such a natural ranking leads to satisfaction of
Assumption~\ref{sigma.A2}.  

According to this assumption, the jammer who seeks a higher value of
payoff should favour channels with lower numbers,
since a larger reward is associated with these channels. In contrast, the
controller actions should be directed towards forcing the jammer into
utilizing channels with higher numbers. Also, the channel 
$f_{j^-}$ is excluded from this ranking. This is done to allow the jammer
to consider contributions to payoff other than those based on
blocking/passing. These considerations may either discourage the jammer
from switching, or conversely encourage it to undertake a 
denial-of-service attack.  
Such decisions can be influenced by a number of factors that are
not related to channel properties. 
The cost of channel switching is one reason as to why the jammer may decide
not to change the channel. Under another scenario, the jammer may be
offered a reward for remaining 
stealthy, and may choose this reward over disrupting the control loop.  
For instance, when the controller monitors the link, an anomaly
in the channel transition probabilities could signal the 
attack. In this case, rewarding the jammer for not defaulting
to the most blocking channel unless it is absolutely necessary will provide
it with an incentive for not revealing itself. In yet another class of
problems, the jammer's decision could be based on the knowledge that the
system is prepared to 
tolerate service disruptions as long as the cost of such disruptions is
below the cost of rectifying them. We defer detailed analyses of these
situations to Section~\ref{Examples}. 
It should be stressed that jammer decisions
in each of these scenarios will depend on the plant state $x$, the
channel $f_{j^-}$ and the channel ranking (the latter may require knowledge
of $u$).  

Using the channel ranking introduced in Assumption~\ref{sigma.A2}, the
value and saddle points of the game (\ref{inf_sup}) can be
characterized by solving a game over a reduced jammer strategy
space. This reduced game focuses on two channels, namely the channel that
currently occupies the link and the channel that delivers the highest
payoff to the jammer when it seeks to block communications between the
controller and the plant. The latter channel is indexed as channel $f_1$, by
Assumption~\ref{sigma.A2}.

Let us introduce the reduced jammer action vector $\tilde p=(\tilde
p_1,\tilde p_2)'$, $\tilde p_1,\tilde p_2\ge 0$,  $\tilde p_1+\tilde
p_2=1$. Also, consider payoffs associated with selecting channel $f_1$ and
keeping the current channel $f_{j^-}$:
\begin{eqnarray*}
\tilde h_1(u)&=&
\mathbb{E}\left[\sigma(x^+,u,f_S)|S=1\right] 
,\\
\tilde h_2(u)&=&
\mathbb{E}\left[\sigma(x^+,u,f_S)|S=j^-\right]
,
\end{eqnarray*}
and define $\tilde h(u)=(\tilde h_1(u),\tilde h_2(u))'$.
Consider the following `reduced' two-player game with upper value
\begin{eqnarray}  
\tilde J_1 &=& \inf_{u} \max_{\tilde p\in \mathcal{S}_1}
\tilde p'\tilde h(u) \label{inf_sup.red}
\end{eqnarray}
and lower value 
\begin{eqnarray}  
\tilde J_2 &=&  \max_{\tilde p\in \mathcal{S}_1} \inf_{u}
\tilde p'\tilde h(u). \label{sup_inf.red}
\end{eqnarray}

\begin{lemma}\label{only.two}
Suppose Assumptions~\ref{sigma.A1}-\ref{sigma.A2} are satisfied. Then 
\begin{eqnarray} \label{JJJJ} 
J_1 =\tilde J_1=\tilde J_2= J_2.
\end{eqnarray}
Furthermore, the zero-sum game (\ref{inf_sup.red}) has a (possibly
non-unique) saddle point. Also, 
if $(u^*,\tilde p^*)$ is such a saddle point, then $(u^*,p^*)$ is a saddle
point of the game (\ref{inf_sup}), where
\begin{equation}
  \label{pp}
  p_j^*=\begin{cases}\tilde p_1^*, & j=1,\\
                     \tilde p_{j^-}^* & j=j^-,\\
                     0                & j\not\in\{1,j^-\}.
                   \end{cases}
\end{equation}
\end{lemma}

\section{Main results}\label{main}

Lemma~\ref{only.two} allows the jammer to constrain its actions
to the set $\bar{\mathcal{S}}=\{p: p_j=0, j\neq 1,j^-\}\subset
\mathcal{S}_{n-1}$. Among these actions there are two trivial actions: choose
the most blocking channel (channel $f_1$ in our notation) by using
$p_1=1$ and $p_j=0, j\neq 1$, or stay put by allocating $p_{j^-}=1$ and
$p_j=0, j\neq j^-$, so that the controller continues communicating with the
plant over channel $f_{j^-}$. However, the question arises as to whether
there exist \emph{optimal mixed strategies} in $\bar{\mathcal{S}}$ for
the jammer to undertake, i.e., optimal policy vectors $p$
such that $0<p_j<1$, $j=1,j^-$.  

The first main result of this paper provides necessary and sufficient
conditions for the existence of nontrivial saddle points. These conditions 
characterize the controller-jammer games in which the jammer randomizes its
choice of optimal strategies. As we will see, this will force the
controller to respond in a non-obvious manner in order to remain optimal, which
ultimately represents a signature of an attack on the communication link. 

\begin{theorem}\label{saddle.point}
Suppose $\tilde h_1$, $\tilde h_2$ are strictly convex functions
of $u$ for all $x$. For every $x$, the zero-sum game (\ref{inf_sup.red})
admits a nontrivial saddle point $(u^*,\tilde p^*)$ if and only if there exists
$\bar u$ such that 
\begin{equation}\label{F=F}
\tilde h_1(\bar u)=\tilde h_2(\bar u),
\end{equation}
and one of the following conditions hold: either
\begin{equation}\label{dF.dF<0}
\left(\frac{\partial\tilde h_1(\bar u)}{\partial u}\right)
\left(\frac{\partial\tilde h_2(\bar u)}{\partial u}\right)<0,
\end{equation}
or
\begin{equation}\label{dF=dF=0}
\frac{\partial\tilde h_1(\bar u)}{\partial u}=
\frac{\partial\tilde h_2(\bar u)}{\partial u}=0.
\end{equation}
\end{theorem}

\section{A linear-quadratic controller-jammer game}\label{Examples}

In this section, we specialize Theorem~\ref{saddle.point} to the
controller-jammer game where the plant is linear,
\begin{equation}
\label{plant.lin}
x^+ = A x + b Bu
\end{equation}
and the performance cost is quadratic. 
In this game the jammer is rewarded for remaining stealthy. We show that in
this game, there is a region in the plant state space where        
the jammer's optimal policy is to randomize its channel selection. Furthermore,
an optimal control response to this optimal jammer action is nonlinear. 

Consider a controller-jammer game for the plant (\ref{plant.lin}) 
with the quadratic payoff 
\[
(\|x\|^2+\|u\|^2)+\|x^+\|^2+ (\delta_{j,j^-})\tau.
\]
Here, $\delta_{j,k}$ is the Kronecker symbol, and $\tau>0$ is the constant
`reward for stealthiness' which the jammer receives if the channel does not
change as a result of its action. As explained earlier, 
the rationale here is to reward the jammer for keeping the current channel
in the link 
when excessive switching may reveal its presence, or may
drain its resources. This controller-jammer game was analyzed
in~\cite{LaU3a} (for a one-dimensional plant), where the region in the
state-space was found where the game has a unique saddle point
corresponding to a jammer's nontrivial strategy. Such a region was found
by computing the game value directly, which required a quite tedious
analysis. Here, we revisit this result of~\cite{LaU3a}  from a more general
perspective, using conditions of Theorem~\ref{saddle.point}.    

The corresponding function $\sigma$ in this case is
\begin{equation}
\sigma(y,u,f_j)=\begin{cases} \|x\|^2+u^2+\|y\|^2, & j\neq j^-, \\
\|x\|^2+u^2+\|y\|^2+\tau,  & j=j^-.
\end{cases}
\label{LQ.cost}
\end{equation}

Clearly, the function $\sigma$ defined in (\ref{LQ.cost})  satisfies
Assumptions~\ref{sigma.A1} and~\ref{sigma.A3}. Also in this case, the
functions $h_j$ have the form 
\begin{eqnarray}
h_j(u)&=&\mathbb{E}[\sigma(x^+,u,f_S)|S=j] \nonumber \\
      &=&\gamma_j(x)+u^2+r_jq_ju(u+2\beta(x)),  
\label{hj.LaU3}
\end{eqnarray}
where 
\begin{eqnarray*}
\gamma_j(x)=\begin{cases} x'(I+A'A)x, & j\neq j^-, \\
 x'(I+A'A)x+\tau, & j=j^-,
\end{cases}
\end{eqnarray*}
$\beta(x)=\frac{1}{\|B\|^2}B'Ax$, and $r_j=\|B\|^2$ for all $j$. Also, the
available channels are assumed to be ordered according to their probability
to become passing, that is, 
\begin{eqnarray}
q_1<q_2<\ldots<q_n.
\label{qq}
\end{eqnarray}

\begin{lemma}\label{U.LQ}
Under condition (\ref{qq}), the set $U(x)=\{u:~u(u+2\beta(x)\le 0\}$ 
verifies properties (i) and (ii) stated in Lemma~\ref{U}, and
also satisfies  Assumptions~\ref{sigma.A4} and~\ref{sigma.A2}. 
\end{lemma}

Under the above assumptions, the payoff functions $\tilde h_1$ and $\tilde h_2$
for the reduced zero-sum game become
\begin{eqnarray}
\tilde h_1(u)= h_1(u) \quad \mbox{and} \quad 
\tilde h_2(u)= h_{j^-}(u).
\label{reduced.LQ}
\end{eqnarray}
With these definitions, condition (\ref{F=F}) reduces to the equation
\begin{equation}
  \label{F=F.LQ}
  \bar u \left(\bar u + \frac{2}{\|B\|^2}B'Ax\right)=
  \frac{\tau}{\|B\|^2(q_1-q_{j^-})}, 
\end{equation}
which admits real solutions if $\frac{1}{\|B\|^2}x'A'BB'Ax\ge
\frac{\tau}{q_{j^-}-q_1}$.    
Also, condition (\ref{dF.dF<0}) reduces to the condition
\begin{equation}
  \label{dF.dF<0.LQ}
  -\frac{\|B\|^2q_{j^-}}{1+\|B\|^2q_{j^-}}< \bar u <
  -\frac{\|B\|^2q_1}{1+\|B\|^2q_1}. 
\end{equation}
Let
\[
z=\frac{\tau \|B\|^2}{(q_{j^-}-q_1)x'A'BB'Ax}.
\] 
The analysis of conditions (\ref{F=F.LQ}), (\ref{dF.dF<0.LQ}) shows that only
one of the solutions of equation (\ref{F=F.LQ}),
\begin{eqnarray}
\bar u&\triangleq &u^* \nonumber \\
&=& -\frac{1}{\|B\|^2}B'Ax\left(1-\sqrt{\frac{\tau
    \|B\|^2}{(q_1-q_{j^-})x'A'BB'Ax}}\right) \quad
\label{u*}
\end{eqnarray}
satisfies (\ref{dF.dF<0.LQ}) provided
\begin{equation}
1-\frac{1}{(1+\|B\|^2q_1)^2} < z < 1-\frac{1}{(1+\|B\|^2q_{j^-})^2}. 
\label{z.LaU3a}
\end{equation}

Condition (\ref{z.LaU3a}) describes the region in the state space in which the
jammer's optimal policy is to choose randomly between the channel $f_{j^-}$
currently in use and the most blocking channel $f_1$. Observe that in the
case where the plant (\ref{plant.lin}) is scalar and $B=1$, we recover
the exactly same condition as that obtained in~\cite{LaU3a} by direct
computation. That is, Theorem~\ref{saddle.point} confirms the existence of
the nontrivial optimal jammer's strategy for this region. We refer the
reader to~\cite{LaU3a} for the exact value of the optimal vector $p^*$;
the calculation for the multidimensional plant (\ref{plant.lin}) follows
the same lines, and is omitted for the sake of brevity. We
also point out that the optimal controller's policy (\ref{u*}) is
nonlinear. Hence, 
any linear feedback policy that controller may employ assuming
that its signals are transmitted over a \emph{bona fide} packet dropping
channel will lead to an inferior control performance. We interpret this
situation as a signature of a successful DoS attack by the jammer.

\section{Discussion and conclusions}\label{Conclusions}

In this paper we have analyzed a class of control problems over
adversarial channels, in which the jammer actively attempts to disrupt
communications between the controller and the plant. We have posed the
problem as a static game, and have given necessary and sufficient
conditions for such a game to have a nontrivial saddle point. The
significance of these conditions is to allow a characterization of a set of
plant's initial states for which a DoS attack can be mounted that requires
a nontrivial controller's response. For instance, in the linear quadratic
problem analyzed in the paper the optimal control law is
nonlinear. This gives the jammer an advantage over any linear control
policy in those problems. The jammer achieves this outcome by randomizing
its choice of a packet-dropping channel rather than operating packet
dropping facility directly. 

On the other hand, the part of the state space
where the jammer randomizes is determined by the jammer's cost of switching
(reward  for not switching) and transition
probabilities of the current and the most blocking channels. If these
parameters can be predicted/estimated by the controller, it has a chance of
mitigating the attack by either eliminating those regions, or steering the
plant so that it avoids visiting those  regions.      

Future work will be directed to further understanding conditions for
DoS attacks, with the aim to consider
dynamic/multi-step control problems. Another interesting question is
whether associating a distinct payoff with one of the channels is
necessary for the jammer to resort to randomization.

\end{document}